\documentclass[11pt,a4paper]{article}
\usepackage[english]{babel}

\usepackage[a4paper, top=2cm, bottom=2cm, left=3cm, right=3cm]{geometry}
\usepackage{setspace}
\usepackage{graphicx}
\usepackage{amsmath, amssymb}
\usepackage{caption}
\usepackage{booktabs}
\usepackage{pifont}
\usepackage{color}
\usepackage{makecell}
\usepackage[colorlinks=true, allcolors=blue]{hyperref}
\usepackage{natbib}
\setcitestyle{authoryear,open={(},close={)}}

\DeclareCaptionLabelFormat{adja-page}{#1 #2 \emph{(previous page)}}
\usepackage{subfig}

\usepackage{lineno}
\usepackage{etoolbox}
\newcommand*\patchAmsMathEnvironmentForLineno[1]{%
  \expandafter\let\csname old#1\expandafter\endcsname\csname #1\endcsname
  \expandafter\let\csname oldend#1\expandafter\endcsname\csname end#1\endcsname
  \renewenvironment{#1}%
    {\linenomath\csname old#1\endcsname}%
    {\csname oldend#1\endcsname\endlinenomath}%
}
\newcommand*\patchBothAmsMathEnvironmentsForLineno[1]{%
  \patchAmsMathEnvironmentForLineno{#1}%
  \patchAmsMathEnvironmentForLineno{#1*}%
}

\patchBothAmsMathEnvironmentsForLineno{equation}
\patchBothAmsMathEnvironmentsForLineno{align}
\patchBothAmsMathEnvironmentsForLineno{multline}
\patchBothAmsMathEnvironmentsForLineno{gather}

\usepackage{listings}
\usepackage{xcolor}
\definecolor{mygreen}{rgb}{0,0.6,0}
\definecolor{mygray}{rgb}{0.5,0.5,0.5}
\definecolor{mymauve}{rgb}{0.58,0,0.82}
\newcommand{\keywords}[1]{%
  \vspace{0.5em}
  \noindent\textbf{Keywords:} #1
}

\title{\textbf{Backcasting biodiversity at high spatiotemporal resolution using flexible site-occupancy models for opportunistically sampled citizen science data}}
\author{}
\date{}

\begin{document}
\maketitle

\begin{abstract}
\normalsize
\noindent For many taxonomic groups, online biodiversity portals used by naturalists and citizen scientists constitute the primary source of distributional information. Over the last decade, site-occupancy models have been advanced as a promising framework to analyse such loosely structured, opportunistically collected datasets. Current approaches often ignore important aspects of the detection process and  do not fully capitalise on the information present in these datasets, leaving opportunities for fine-grained spatiotemporal backcasting untouched. We propose a flexible Bayesian spatiotemporal site-occupancy model that aims to mimic the data-generating process that underlies common citizen science datasets sourced from public biodiversity portals, and yields rich biological output. We illustrate the use of the model to a dataset containing over 3M butterfly records in Belgium, collected through the citizen science data portal Observations.be. We show that the proposed approach enables retrospective predictions on the occupancy of species through time and space at high resolution, as well as inference on inter-annual distributional trends, range dynamics, habitat preferences, phenological patterns, detection patterns and observer heterogeneity. The proposed model can be used to increase the value of opportunistically collected data by naturalists and citizen scientists, and can aid the understanding of spatiotemporal dynamics of species for which rigorously collected data are absent or too costly to collect.
\end{abstract}


\section*{Introduction}\label{introduction}

Citizen science initiatives are rapidly gaining popularity around the
globe. Online platforms such as eBird, iNaturalist and Observation.org
enable volunteer naturalists to easily report sightings of wild species,
generating data on species occurrences and abundances at an
unprecedented rate. Such citizen science records have proven to be a
valuable and practical resource for conservationists, scientists and
policy-makers, mainly due to their retrospective availability and their
spatial, temporal and taxonomic extent and detail (Maes \emph{et al.} 2015;
Robinson \emph{et al.} 2018; Soroye \emph{et al.} 2018).

Despite these attractive properties, citizen science data generally come
with a multitude of shortcomings and challenges, impeding
straightforward analysis. For instance, even though they cover vast
geographical areas, extensive time windows and a wide taxonomic scope,
citizen science datasets are typically sparse, leaving many gaps when
considering individual locations, time frames and taxa (Outhwaite
\emph{et al.} 2018; Pocock \emph{et al.} 2019). Opportunistic sampling
schemes (in which participants can freely choose where, when and which
species to sample) also tend to suffer from an unequal distribution of
sampling effort through space, time and taxa (Johnston \emph{et al.}
2020; Neyens \emph{et al.} 2019; Ruiz-Gutierrez \emph{et al.} 2016), with sampling concentrated in natural areas and parks in densely populated regions, or around cities in more sparsely inhabited regions.
While the growing popularity of citizen science initiatives tends to
partially remediate data sparseness, imbalance can be expected to grow
due to a broadening audience (Shirey \emph{et al.} 2020). Citizen
science portals tend to target and harbour a large heterogeneity among
observers (Fitzpatrick \emph{et al.} 2009; Johnston \emph{et al.} 2018),
ranging from highly skilled naturalists to novice recorders starting out
with automatic species recognition apps such as ObsIdentify, Merlin and
Pl@ntNet (Hogeweg \emph{et al.} 2019; Joly \emph{et al.} 2016).
Increasingly heterogeneous user bases also tend to increase the
probability of false positive records, e.g. through species
misidentifications (Ferguson \emph{et al.} 2015; McClintock \emph{et
al.} 2010; Miller \emph{et al.} 2011; Royle \& Link 2006). Despite
extensive quality control, such errors may persist and can further
obfuscate biological patterns inferred from citizen science data
(Vantieghem \emph{et al.} 2017). Identification challenges might also prevent sufficiently careful observers to report observations, further inflating imperfect detection and reducing detection rates, regardless of the actual occupancy or abundance of species. Consequently, "occupancy" should be interpreted as a site's state of having been occupied at least once throughout the year.

Virtually all types of biodiversity surveys, including opportunistic sampling schemes, are prone to false negatives (Kellner \& Swihart 2014).
False negatives typically arise from imperfect detection, i.e. the act
of not detecting an individual or species at a location even though it
is present (Kéry \& Schmidt 2008; MacKenzie \emph{et al.} 2002). In opportunistic sampling schemes in particular, imperfect detection can assume strongly heterogeneous forms, varying across observers, visit-specific circumstances and throughout the year (Kellner \& Swihart 2014). Moreover, citizen science initiatives also suffer from reporting bias,
as individuals or species might not always be reported even though the observer detected them (Van Strien \emph{et al.} 2013). Such reporting bias also leads to false negatives and will therefore manifest itself in the same way imperfect detection does.

MacKenzie et al. (2002) developed the site-occupancy framework to
address imperfect detection, by means of a hierarchical modelling
approach that disentangles the observation process from the latent
biological process through repeated site visits. Site-occupancy models
have quickly gained popularity, mainly in wildlife survey studies. Over
the past decade, site-occupancy models have been proposed as a promising
framework to analyse opportunistically sampled data (Kéry \emph{et al.}
2010; Van Strien \emph{et al.} 2013).
Several studies have demonstrated and validated the application of
site-occupancy to citizen science data (Burns \emph{et al.} 2018;
Outhwaite \emph{et al.} 2018, 2020; Van Strien \emph{et al.} 2013, 2016,
2019; Termaat \emph{et al.} 2015, 2019). The type of considered datasets
range from semi-structured checklist data to loosely structured
opportunistic data. In the present study, we focus on the latter
category, as it is the broadest type of biodiversity data and is more
challenging to deal with. Such data arises whenever naturalists record
and submit sightings of wild species to a data portal, without the need
of a protocol dictating when, where, how or what to record \emph{a
priori}, except that the sighting should be temporally and spatially
referenced. Typically, pseudo-visits are inferred from loosely
structured opportunistic data based on reported observations of species
other than the focal species (Van Strien \emph{et al.} 2013). In many
studies, list length, i.e. the number of recorded species during such a
pseudo-visit, has been used as a covariate to model detection
probability and serves as a proxy for search effort (Isaac \emph{et al.}
2014; Outhwaite \emph{et al.} 2018; Van Strien \emph{et al.} 2013).
Seasonal patterns in detection probability are often accommodated with
ad hoc approaches, such as restricting the time window of the used input
data, and ad hoc approaches are also used to address spatial variation
in occupancy, such as only considering sites with at least one detection
of the focal species over the entire study period (Van Strien \emph{et
al.} 2013). Interannual variation in occupancy is generally modelled by means of a
dynamic site-occupancy specification (Van Strien \emph{et al.} 2011) or
through more efficient longitudinal structures (Outhwaite \emph{et al.}
2018).

Recently, Belmont \emph{et al.} (2024) demonstrated how spatiotemporal
structure can be incorporated into site-occupancy models to facilitate
information exchange across space and time, thereby capturing
biologically realistic, temporally varying distributional patterns.
Their approach uses Integrated Nested Laplace Approximation (INLA),
which substantially reduces computational burden compared to Bayesian
approaches relying on sampling (Belmont \emph{et al.} 2024). However, the implementation in INLA only allows for limited opportunities to model variation in detectability at the visit level, requiring the assumption of a  simplified detection process. For opportunistically collected biodiversity portal
data in particular, this assumption might be too restrictive given the
highly heterogeneous and imperfect detection process, and it is unclear
whether the computational gains justify the simplification.

In the present study, we propose a spatiotemporal site-occupancy model
for opportunistically collected biodiversity portal data, that flexibly
models the detection process to retrospectively predict occupancy
patterns at high resolution across space and time. By capitalising on
modern advances in hierarchical modelling, the approach mimics the data
collection process as closely as possible, in order to counter important
biases and to efficiently exploit information hiding in the data. As a
case study, we apply our model to a dataset of butterfly (Lepidoptera:
Rhopalocera) sightings in Belgium (Vanreusel \emph{et al.} 2019),
collected through \href{https://waarnemingen.be/}{Observations.be} (`Waarnemingen.be' in Dutch), the
Belgian branch of the Observation.org biodiversity data portal, managed by Natuurpunt Studie and Natagora. Observations.be contains one of the highest biodiversity record
densities in the world (on average 338 records per km² in the year 2024,
across all taxa). Butterflies are a well-studied group of species and strong declines have been observed over the past century in Western
Europe (Warren \emph{et al.} 2021) and in Belgium in particular (Maes \&
Van Dyck 2001), highlighting the need for an improved understanding of
spatiotemporal patterns.

\section*{Methods}\label{methods}

\subsection*{Data preparation
procedure}\label{data-preparation-procedure}

We derive discrete pseudo-visits from a database of opportunistically sampled,
temporally and spatially referenced, multi-taxon sightings by
naturalists. Due to the loose nature of these records, no formal
information is available on the precise sampling behaviour and visit
metadata of the observers. Hence, we use the pseudo-visit strategy
proposed by Van Strien \emph{et al.} (2013) instead: a visit \(v\) is
defined whenever a particular observer reports at least one sighting of
any species belonging to the focal taxon on a particular day in a particular site. Sites are defined
as individual cells in a grid defined over the study area. The outcome
variable \(y_{v}\) indicates whether the focal species has been reported
(\(y_{v} = 1\)) during visit \(v\) or not (\(y_{v} = 0\)). For each
visit, additional metadata comprises the visited site's ID \(s(v)\), the
year \(t(v)\), the observer ID \(o(v)\), the week \(w(v)\ \)and a vector
of additional detection-related covariates \(X_{v}^{p}\). In the
following, we only consider list length \(l_{v}\) (the number of
recorded species belonging to the focal taxon during visit \(v\)) as a detection-related covariate
(categorised: 1 species, 2-3 species, \textgreater{} 3 species;
following Van Strien et al. 2013), as it can easily be derived from the
visit data and because it has successfully been included in prior
studies (Van Strien \emph{et al.} 2013), though extending the suite of
considered detection-related covariates (e.g. by better proxies of
search effort if available, daily meteorological variables, \ldots) is
trivial in our model. A second dataset comprises a vector of
occupancy-related covariates \(X_{s}^{\psi}\) for each of the sites
\(s\). To account for spatial dependencies, the spatial coordinates
\(\text{lon}(s)\) and \(\text{lat}(s)\) of the grid cell centroids are
also computed. For each site and year combination, the binary variable
\(a_{s,t}\) represents whether or not the presence of the species has
been confirmed (i.e. at least one positive detection or any other source
of evidence).

\subsection*{Core model structure}\label{core-model-structure}

Following the site-occupancy modelling approach (MacKenzie \emph{et al.}
2002), our model aims to disentangle the detection process from the
latent biological process. Hence, the observed detection outcome
\(y_{v}\) is assumed to follow a Bernoulli distribution:

\[y_{v}\ |\ z_{s(v),t(v)} \sim \operatorname{Bernoulli}\left( p_{v} \cdot z_{s(v),t(v)} \right),\]

where \(p_{v}\) is the detection probability during visit \(v\) and
\(z_{s(v),t(v)}\) is the latent occupancy status of the visited site
\(s\) during year \(t\). The latent occupancy status \(z_{s,t}\) of site
\(s\) during year \(t\) is also assumed to follow a Bernoulli
distribution:

\[z_{s,t} \sim \operatorname{Bernoulli}\left( \psi_{s,t} \right),\]

where \(\psi_{s,t}\) is the occupancy probability of site \(s\) during
year \(t\). For identifiability, the occupancy status of a site is assumed to only vary across years and to be constant within a given year (i.e. the 'closure assumption', MacKenzie \emph{et al.} 2002). Phenological unavailability (e.g. undetectable or intentionally discarded developmental stages) or temporary immigration or emigration within a year are captured by the detection part of the site-occupancy model (cf. infra).

For efficiency reasons, and due to computational limitations
regarding latent discrete variables, we marginalise out the latent
occupancies (Yackulic \emph{et al.} 2020), leading to the following
two-case likelihood structure which depends on \(a_{s,t}\), i.e. whether
the presence of the focal species has been confirmed for site \(s\)
during year \(t\):

\[L\left( \mathbf{y}_{s,t} \right) = \left\{ \begin{matrix}
\psi_{s,t}\prod_{v \in \mathbf{V}_{\mathbf{s,t}}}^{}{{p_{v}}^{y_{v}}\left( 1 - p_{v} \right)^{1 - y_{v}}}, & \text{if}\ a_{s,t} = 1 \\
\left( 1 - \psi_{s,t} \right) + \psi_{s,t}\prod_{v \in \mathbf{V}_{\mathbf{s,t}}}^{}\left( 1 - p_{v} \right), & \text{if}\ a_{s,t} = 0,
\end{matrix} \right.\ \]

where \(\mathbf{y}_{s,t}\) constitutes the detection outcomes during the
set of visits \(\mathbf{V}_{\mathbf{s,t}}\) performed at site \(s\)
during year \(t\). The simplest possible site-occupancy model only
features intercepts, where the detection probabilities \(p_{v}\) and the
occupancy probabilities \(\psi_{s,t}\) are assumed to be constant
throughout space and time (\(p_{v} = \beta_{0}^{p}\),
\(\psi_{s,t} = \beta_{0}^{\psi}\)). This stringent assumption can be
relaxed by modelling the detection probabilities and occupancy
probabilities through the logistic regression framework, which is the
approach we take. More specifically, we will outline several extensions
to the intercept-only model in the following sections to achieve a
flexible model that acknowledges the typical biases and complexities of
opportunistically sampled citizen science data.

\subsection*{Modelling detection
probabilities}\label{modelling-detection-probabilities}

In our model, we assume that the probability of detecting the focal
species during a visit depends on the intrinsic detectability of the
species, on the performed search effort, on the propensity of the
observer to detect and to report the focal species and on phenological
detectability patterns, which may vary seasonally. Specifically, we
model the detection probability \(p_{v}\) during visit \(v\) as follows,
on the logit scale:

\[\log\left( \frac{p_{v}}{1 - p_{v}} \right) = \ \beta_{0}^{p} + X_{v}^{p}\beta^{p} + \ b_{o(v)}^{obs} + \ f_{phen}\left( w(v) \right),\]

where \(\beta_{0}^{p}\) is the overall detection probability intercept
(representing the average detectability of a species on the logit
scale), \(\beta^{p}\) is a vector of regression coefficients
representing the effects of the detection-level covariates,
\(b_{o}^{obs}\) is a normally distributed individual random intercept for
observer \(o\), and \(f_{phen}\) is a smooth function representing the
seasonal phenological variation, evaluated at week \(w\). We model
\(f_{phen}\) using a Gaussian process (GP) prior with a periodic
covariance function (Rasmussen \& Williams 2006):

\[f_{phen\ }\sim\ \mathcal{GP}\left( 0,k_{phen} \right)\ \ \ \text{and}\ \ \ k_{phen}\left( w,w' \right) = \sigma_{phen}^{2}\exp\left( - \frac{2\text{ sin}^{2}\left( \pi\frac{\left| w - w' \right|}{53} \right)}{l_{phen}^{2}} \right),\]

where \(\sigma_{phen}\) is a marginal standard deviation parameter,
\(l_{phen}\) is a length scale parameter, and 53, the total number of
distinct weeks in a year, is the periodicity. Gaussian processes are a
powerful approach to model time series or latent continuous functions in
general, as the length-scale parameter tunes the smoothness of the curve
based on the data, and determines how rapidly the function can change
(e.g. a smooth versus a rapidly varying phenological curve). The use of a periodic covariance function ensures that seasonal patterns are continuous on an annual basis.

In addition to the phenological term, visit-specific meteorological predictors could easily be supplied as part of the detection-level design matrix $X^{p}$. Even though meteorological conditions can affect detection probabilities, we did not include them as predictors because our goal was to characterize the realized phenology as observed in the field, regardless of its underlying drivers. Including meteorological variables would have blocked part of the genuine phenological signal—namely the proximate, weather-induced variation that forms an integral component of the realized activity pattern. Additionally, most observers tend to avoid unfavourable meteorological conditions to perform visits, and poor meteorological conditions typically lead to empty species lists, precluding these visits from being modelled. As a result, true meteorological patterns might be difficult to capture.

\subsection*{Modelling occupancy
probabilities}\label{modelling-occupancy-probabilities}

We assume that the probability that a site is occupied by the focal
species depends on the overall widespreadness of the species, the effect
of a set of occupancy-related covariates as well as on the temporal,
spatial and spatiotemporal context. Specifically, we model the occupancy
probability \(\psi_{s,t}\) of site \(s\) during year \(t\) as follows,
on the logit scale:

\[\log\left( \frac{\psi_{s,t}}{1 - \psi_{s,t}} \right) = \beta_{0}^{\psi} + \ X_{s}^{\psi}\beta^{\psi} + \delta_{t} + \upsilon_{s} + \varsigma_{s}t^{*},\]

where \(\beta_{0}^{\psi}\) is the overall occupancy probability
intercept (representing the average occupancy of a species on the logit
scale), \(\beta^{\psi}\) is a vector of regression coefficients
representing the effects of the occupancy-level covariates,
\(\delta_{t}\) represents the temporal random effect for year \(t\),
\(\upsilon_{s}\) is a spatial random effect for site \(s\) and
\(\varsigma_{s}\) is a spatial random linear trend slope for site \(s\),
which captures local, linear deviations from the overall trend across
space (Knorr-Held 2000). The spatial linear trend slope
\(\varsigma_{s}\) is multiplied with the year \(t^{*}\), scaled on the
\(\lbrack - 0.5,0.5\rbrack\)-interval for interpretability reasons.

The temporal trend values \(\delta_{t}\) are modelled as follows:

\[\delta_{t} = \beta^{\delta}t^{*} + \delta_{t}^{GP} + + \delta_{t}^{i.i.d.},\]

where \(\beta^{\delta}\) is a linear trend slope to capture any
directional trend, \(\delta_{t}^{GP}\) is a smooth Gaussian process
component to flexibly model temporally autocorrelated patterns, and
\(\delta_{t}^{i.i.d.}\) is a normally distributed yearly noise term. We
use a Gaussian process prior with an exponentiated quadratic covariance
function to model \(\delta^{GP}\):

\[\delta^{GP}\ \sim\ \mathcal{GP}\left( 0,k_{\delta} \right)\ \ \ \text{and}\ \ \ k_{\delta}(t,t') = \sigma_{\delta}^{2}\exp\left( - \frac{\left( t - t' \right)}{2l_{\delta}^{2}}^{2} \right),\]

where \(\sigma_{\delta}\) is a marginal standard deviation parameter,
\(l_{\delta}\) is a length scale parameter, dictating the rate at which
the non-linear trend changes.

The spatial random effects \(\upsilon_{s}\) for each site \(s\) are
modelled as a linear combination of spatially correlated (i.e.
structured) and uncorrelated (i.e. unstructured) random effects,
weighted by the spatial signal \(p\):

\[\upsilon_{s} = (1 - p)\upsilon_{s}^{unstr} + p\upsilon_{s}^{str}\]

where \(\upsilon_{s}^{unstr}\) and \(\upsilon_{s}^{str}\) represent the
spatially unstructured and structured random effects respectively. The spatial signal $p$ defines the relative importance between the spatially structured and unstructured terms, and allows the estimation of a single scale parameter $\sigma^{space}$. The spatially unstructured term captures variation in occupancy probability among sites that varies noisily, but it can also capture spatially structured variation for which the spatially structured term is too coarse. The
approach and description is analogous to the one used in Fajgenblat \&
Neyens (2025). The spatially unstructured random effects are assumed to
be normally distributed:

\[\upsilon_{s}^{unstr}\sim\ \text{Normal}\left( 0,\sigma^{space} \right),\]

with \(\sigma^{space}\) a scale parameter. Similarly to the temporal
random effects, we use Gaussian processes to model the spatially
structured random effects \(\upsilon_{s}^{str}\). Exact Gaussian
processes become impractical when being evaluated over more than a
couple of hundred input locations due to their cubically scaling
computational complexity. We use B-splines projected Gaussian processes
(Monod \emph{et al.} 2022) instead to facilitate the evaluation over a
larger number of locations. B-splines projected Gaussian processes are
similar to penalised splines but have been shown to outperform them
(Monod \emph{et al.} 2022). First, we define a two-dimensional
tensor-product spline surface with \(n\) basis functions along each
dimension over the study area\textquotesingle s square bounding box:

\[\upsilon_{s}^{str} = \sum_{g = 1}^{n}{\sum_{h = 1}^{n}\left( w_{g,h} \cdot b_{g}\left( \text{lon}(s) \right) \cdot b_{h}\left( \text{lat}(s) \right) \right)},\]

where \(\text{lon}(s)\) and \(\text{lat}(s)\) are the longitude and
latitude of site \(s\), \(b_{g}( \cdot )\) and \(b_{h}( \cdot )\)
are the \(g\)'th and \(h\)\textquotesingle th cubic basis functions
anchored to equally spaced knots along each dimension, and \(w_{g,h}\)
is their corresponding weight coefficient. Since most study areas do not
match an exact square, some basis function products might equal zero (or
some negligible value) across the entire study area. Accordingly, these
basis functions combinations do not need to be evaluated and the
corresponding weight coefficients \(w_{g,h}\) do not need to be
estimated, easing computation. The full set of B-spline weight
coefficients \(\mathbf{w}\) are modelled through an exact Gaussian
process:

\[\mathbf{w}\mathcal{\ \sim\ GP}\left( 0,k_{w} \right)\ \ \ \text{and}\ \ \ k_{w}\left( (g,h),\left( g',h' \right) \right) = \sigma_{w}^{2}\exp\left( - \frac{\left( g - g' \right)^{2} + \left( h - h' \right)^{2}}{2l_{w}^{2}} \right),\]

where \(k_{w}\) is an exponentiated quadratic covariance function that
dictates how the covariance between two weight coefficients decays as a
function of the Euclidean distance between the basis function indices,
with the marginal standard deviation parameter \(\sigma_{w}\) and the
length scale parameter \(l_{w}\). The spatial random linear trend slopes
\(\varsigma_{s}\) are assumed to be spatially smooth, and are modelled
in an identical fashion as the spatially structured random effects
\(\upsilon_{s}^{str}\). If deemed appropriate, more complex
spatiotemporal model components, such as an anisotropic,
three-dimensional Gaussian process can be used instead.

\subsection*{Controlling for false
positives}\label{controlling-for-false-positives}

The likelihood structure (cf. section "Core model structure" implicitly assumes
false positive detections to be absent. As violations of this assumption
have the potential to strongly bias occupancy estimates, false positive
detections deserve particular care, especially when dealing with data
prone to such errors, such as opportunistically collected data. The
\emph{confirmed presence model} presented by (Ferguson \emph{et al.}
2015) distinguishes positive detections that are uncertain (e.g. by
unexperienced observers) and certain (e.g. by experienced observers,
with photographic proof \ldots), which, along with non-detections, are
treated as realisations of a categorical distribution. In addition to
detection and occupancy probabilities, probabilities of generating false
positive detections are estimated as part of these models. Within the
context of opportunistically collected data, these probabilities can be
assumed to be strongly heterogeneous across observers and through time,
requiring an additional set of linear predictors to be modelled. While
potentially powerful, these modifications to the ordinary site-occupancy
model strongly increase computational complexity, which renders this
approach unpractical for large projects.

Instead, we propose an alternative approach relying on data filtering (Van Eupen \emph{et al.} 2021),
by creating two data streams of which the respective strengths are
harnessed appropriately. More specifically, observers matching objective
or subjective criteria (e.g. having recorded a specific number of
different species of the focal taxon) are considered to be proficient, and are assumed to
display a negligibly small probability of generating false positive
detections. Non-detection visits by these observers are also assumed to
be informative on the absence of the focal species. Hence, visit data of
these observers are used to feed the actual site-occupancy model
likelihood. Nevertheless, other observers also have the potential to
collect useful ecological information that would otherwise be discarded
if their records were to be completely omitted, especially if records
have been validated through photographic or circumstantial evidence by
expert data reviewers. The above-mentioned marginalisation of the latent
occupancy statuses offers a convenient way of still including such
records as the site- and year-specific likelihood is broken up into two
scenarios depending on whether the focal species has been confirmed
(\(a_{s,t} = 1\)) or not (\(a_{s,t} = 0\)) for each site and year
combination. In our approach, both records originating from proficient
observers as well as validated records from other observers are used to
establish \(a_{s,j}\). As such, validated records can still bring
valuable information, even though they were collected by non-proficient
observers. Note that this approach can also be used to include
information on the confirmed presence of the focal species from any
trustworthy source (e.g. distributional atlases), offering opportunities
for data fusion.

\subsection*{Bayesian inference through Hamiltonian Monte
Carlo}\label{bayesian-inference-through-hamiltonian-monte-carlo}

We implemented the model using the probabilistic programming language
Stan (Carpenter \emph{et al.} 2017). Stan performs Bayesian inference by
means of dynamic Hamiltonian Monte Carlo (HMC), a gradient-based Markov
chain Monte Carlo (MCMC) sampler (Betancourt 2017). In general, HMC
outperforms other MCMC algorithms such as Gibbs sampling or
Metropolis-Hastings for highly dimensional models, such as ours
(Monnahan \emph{et al.} 2017; Yackulic \emph{et al.} 2020).

We use the `CmdStanR' package as an interface to Stan v2.36.0, in R
v4.0.3 (R Core Team 2020). We ran eight chains using 1,000 iterations
each, of which the first 500 were discarded as warm-up, yielding 4,000
posterior samples.

We assessed model convergence both visually by means of trace plots (for
a random subset of parameters) and numerically by means of effective
sample sizes, divergent transitions and the Potential Scale Reduction
Factor, for which all runs had \(\widehat{R} < 1.1\) (Vehtari \emph{et
al.} 2019). In addition, we screened HMC diagnostics (e.g. absence of
divergent transitions).

\subsection*{Prior specifications}\label{prior-specifications}

We assume vaguely informative \(\operatorname{Normal}(0,3)\) priors on detection
intercept and regression coefficients, on the occupancy intercept and
regression coefficients, and on the interannual slope. We assume vaguely
informative \(\operatorname{Normal}^{+}(0,3)\) priors on the scale parameters of
the observer random effects, phenological Gaussian process, the
temporally unstructured random effects, the temporal Gaussian process,
the spatial random effects, and the spatial random trend slopes. We
assume a mildly informative \(\operatorname{InvGamma}(5,5)\) priors on all
included Gaussian processes. By scaling the input values of these
Gaussian processes on the interval \(\lbrack - 1,1\rbrack\), this prior
favours reasonable length scales, avoiding unlikely wiggly or smooth
curves. We assume a flat \(\operatorname{Uniform}(0,1)\) prior on the spatial
signal \(p\). All hierarchical structures rely on a non-centred
parametrisation, with a zero-sum constraint on the standard normal
distributions to improve parameter identifiability. This constraint is
implemented using Stan's efficient ``sum\_to\_zero\_vector'' variable
type, which relies on an inverse logarithmic ratio transform (Carpenter
et al. 2017). To ensure standard normality, we use a
\(\operatorname{Normal}\left( 0,\sqrt{\frac{N}{N - 1}} \right)\) prior in
combination with the ``sum\_to\_zero\_vector'' variable type, with \(N\)
the number of vector elements.

\subsection*{Application: butterflies in
Belgium}\label{application-butterflies-in-belgium}

We applied the outlined approach to butterflies in Belgium as a case
study. Visits and detection/non-detection data are derived from raw
sightings recorded on the Belgian data portal Observations.be between
01/01/2009 and 31/12/2024, by linking each sighting to its corresponding
1 x 1 km grid cell. Additionally, visit metadata comprising the site ID,
year, (anonymised) observer ID, week of the year and list length is
constructed. Observers having submitted at least 500 butterfly sightings
are heuristically deemed proficient and we assume their probability of
generating false positive observations to be negligibly small. Since these observers have collectively performed most visits, this data filtering step only leads to a minor decrease in total data size. Moreover, data from
all other observers, which has been validated by experts through the
availability of photographic or circumstantial evidence, contribute to
establishing whether the species has been confirmed for a given site and
year (\(a_{s,t} = 1\)). Only sightings of living, adult butterflies were
considered as positive detections, while all sightings (including dead
individuals and immature stages) contribute to establishing whether the
species has been confirmed for a given site and year.

Since Observations.be is a multi-taxon portal, the number of recorded
butterfly species during a visit (i.e. list length) can equal zero (e.g.
a birding observer in winter). To reduce computational burden, we
omitted all visits with a zero list length for the focal taxonomic group from our analysis, thereby
strongly reducing the number of visits and speeding up model estimation.
As a drawback, detection probabilities should be interpreted as the
probability of detecting the focal species during a visit, given that at
least one butterfly species has been sighted during the visit. This
particularly affects species that are phenologically active when few
other species are (i.e. species wintering as adults), leading to
inflated estimates of phenological patterns.

With respect to the occupancy-related covariates, we consider eight land
cover variables derived from the Corine Land Cover plus Backbone 2018
(European Environment Agency 2022) and two topographic variables derived
from a 20 x 20 m digital terrain model (NGI 2022). The land cover
variables comprise the classes ``sealed'', ``woody -- needle leaved
trees'' (henceforth abbreviated as ``needleleaved trees''), ``woody --
broadleaved deciduous trees'' (abbr. ``broadleaved trees''),
``low-growing woody plants (bushes, shrubs)'' (abbr. ``shrubland''),
``permanent herbaceous'' (abbr. ``permanent grassland''), ``periodically
herbaceous'' (abbr. ``periodic grassland''), ``water'', ``non- and
sparsely-vegetated'' (abbr. ``bare ground''), and are expressed as
fractions ranging from \(0\) to \(1\). Due to compositionality of land
cover classes (i.e. all land cover fractions sum to one), the dominant class (periodic grassland) is not included as a covariate. In visualisations, however, the effect of varying
fractions for all eight land cover classes is shown by evaluating
posterior predictions over the range from 0 to 1, while proportionally
increasing or decreasing the fractions of other land cover variables. Since we only consider static (shapshot) land cover data, any land cover change-induced changes occupancy probability will be captured through the temporal or spatiotemporal effects included in the model. The two topographic variables are the elevation mean and standard deviation of each grid cell, which are also scaled
between 0 and 1 but back-transformed to their original scale for
visualisation. We use 20 x 20 basis functions for the 2D spatial and
spatiotemporal B-spline projected Gaussian processes.

\section*{Results}\label{results}

Output for one example species, the Purple emperor \emph{Apatura iris},
obtained by applying our modelling approach, is illustrated through
Figures 1-3. Detection probabilities vary strongly across visits (Figure
1). The probability of detecting a species depends on the list length,
our proxy for search effort, and is substantially higher for list
lengths exceeding three species (Figure 1a). Additionally, even when accounting for the effect of list length, we observe a
strong heterogeneity among observers, with average posterior mean
detection probabilities ranging from 1.6\% to 46.0\% (Figure 1b).
Detectability shows a strong seasonal pattern, with a peak during the
week of June 27 (Figure 1c).

\begin{figure}
    \centering
    \includegraphics[width=1\linewidth]{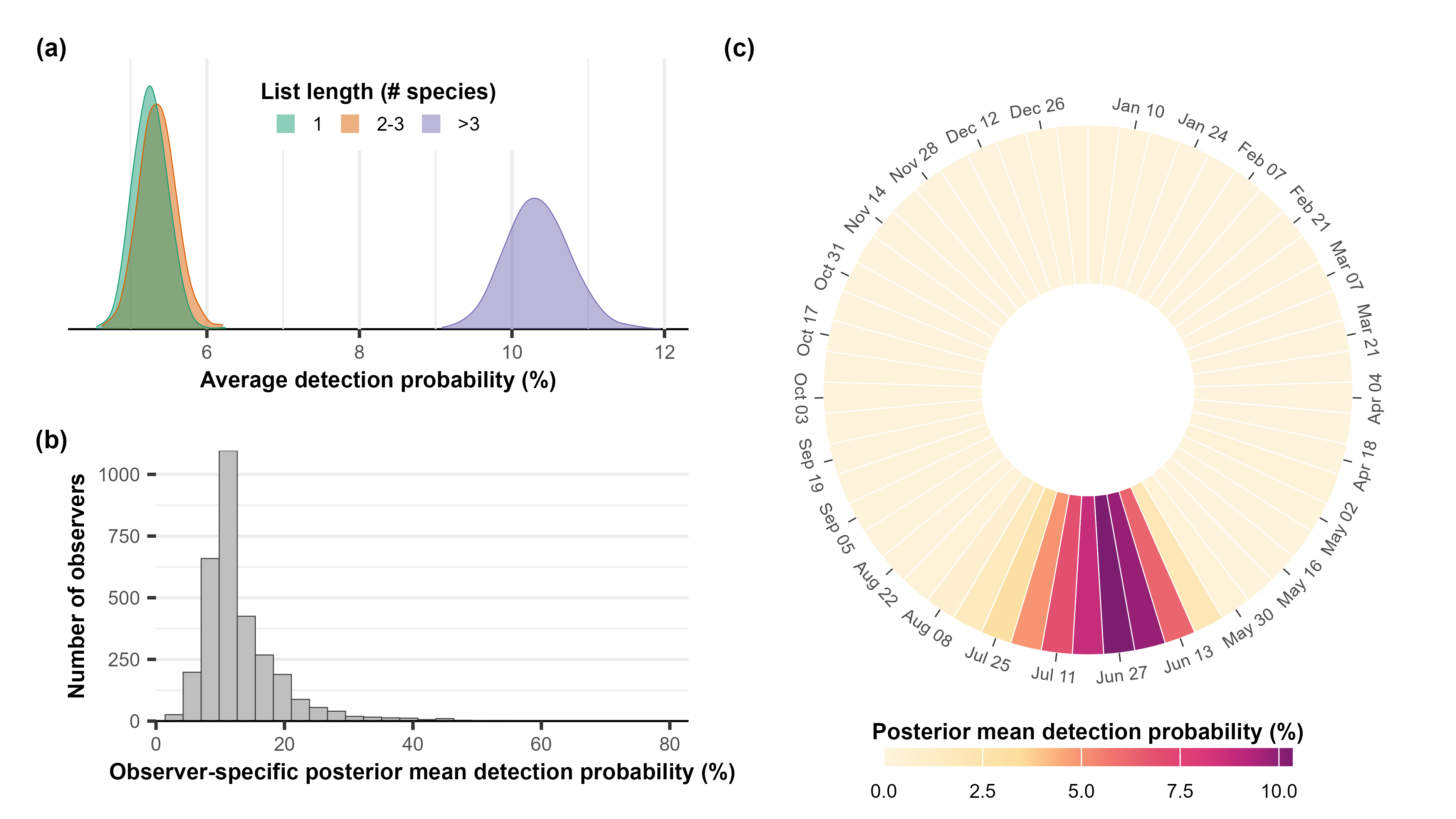}
    \caption{Detection-related output for one example butterfly
species, \textit{Apatura iris}. (a) Posterior detection probabilities for
the three list length classes, for an average observer during the
phenological peak activity week. (b) Histogram of posterior mean
observer-specific detection probabilities, given a list length
\textgreater{} 3 during the phenological peak activity week. (c)
Posterior mean weekly detection probabilities, given a list length
\textgreater{} 3 and an average observer.}
    \label{fig:1}
\end{figure}

The model produces a posterior distribution on the occupancy probability
for each site and year combination, which can be summarised using
posterior means and visualised as yearly distributional maps (Figure 2),
revealing spatiotemporal patterns in the distribution of the focal
species. For the example species \emph{Apatura iris}, a strong expansion
can be appreciated throughout the study period, particularly in Northern
Belgium, where the species' distribution was formerly restricted to a
few regions (Figure 2, Figure 3a). The spatial random trend slopes confirm that
occupancy probabilities have strongly increased in the northern part of
the country throughout the study period, while they were stable or
slightly negative in other parts (Figure 3b). By averaging the posterior
occupancies across space per year, distributional trends are obtained
(Figure 3c), revealing changes in the fraction of the study area
occupied by the focal species. For the example species \emph{Apatura iris}, strong distributional expansions can be observed between 2016-2018 and again between 2021-2022. Interannual distributional trends can be derived
for any subregion by selecting the corresponding sites. For instance,
the distributional trends at the provincial level reveal that the
strongest distributional increase took place in the province of Antwerp
(Figure 3d). Finally, the model produces inference on associations with
distributional covariates (Figure 3e). With respect to land use, the
example species \emph{Apatura iris} shows the strongest positive
association with broadleaved trees and water as land cover. We also
observe high posterior support (\textgreater{} 95\% posterior
probability) for a positive association with both elevation mean and
elevation standard deviation.

\begin{figure}
    \centering
    \includegraphics[width=1\linewidth]{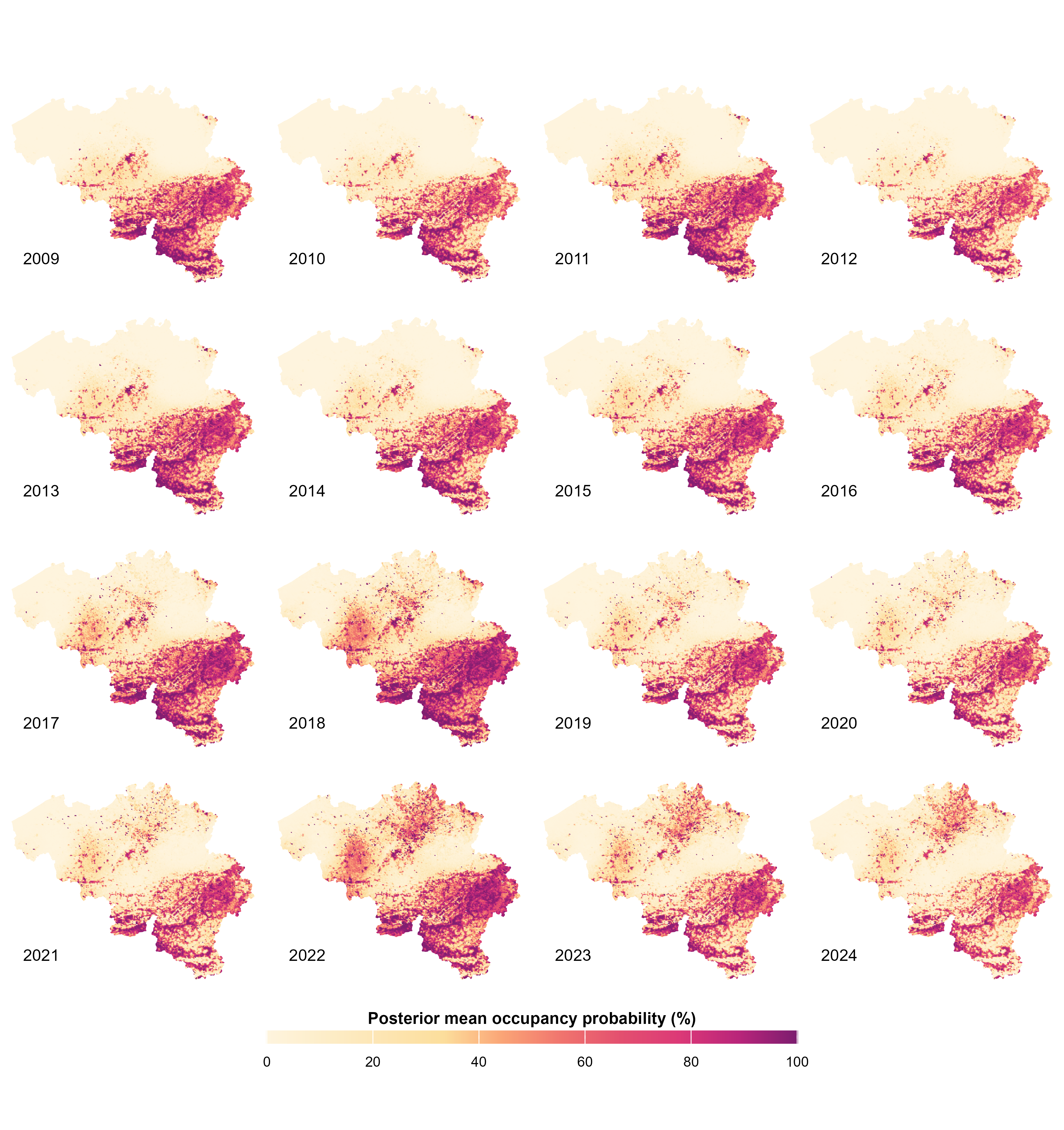}
    \caption{Predicted distribution of one example butterfly species, \textit{Apatura iris} in Belgium across the years 2009-2024. The colours reflect posterior mean occupancy probabilities, with darker colours corresponding with higher probability of occupancy.}
    \label{fig:2}
\end{figure}

\begin{figure}
    \centering
    \includegraphics[width=1\linewidth]{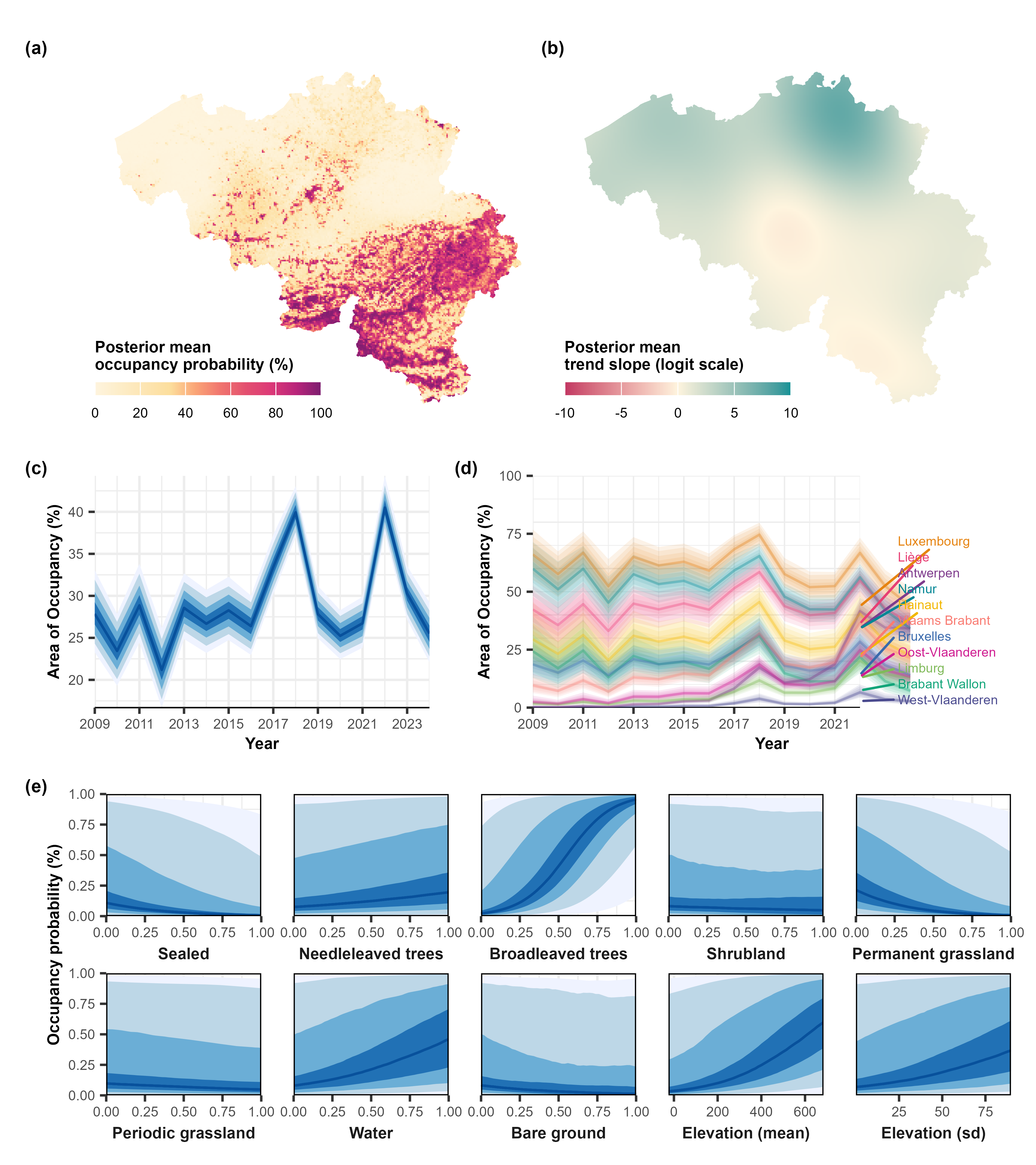}
    \caption{Occupancy-related output for one example butterfly species, \textit{Apatura iris}. (a) Posterior mean occupancy patterns, averaged of the years of the study period. (b) Posterior mean interannual trend slopes. (c) Temporal trend of the fraction of occupied 1 x 1 km grid cells in Belgium. The solid line represents the posterior mean, while the colour gradient reflects the posterior uncertainty (50, 80, 95 and 99\% credible intervals, with increasing transparency). (d) Temporal trend of the fraction of occupied 1 x 1 km grid cells in each Belgian province. The solid lines represent the posterior means, while the colour gradients reflect the posterior uncertainty (50, 80, 95 and 99\% credible intervals, with increasing transparency), with different colours per province. (e) Marginal effects of the included environmental variables on occupancy. The x-axis shows the values of the environmental variables, and the y-axis shows the expected occupancy probability, for the average grid cell and year. The solid line represents the posterior mean, while the colour gradient reflects the posterior uncertainty (50,
80, 95 and 99\% credible intervals, with increasing transparency).}
    \label{fig:3}
\end{figure}

\section*{Discussion}\label{discussion}

By disentangling the detection process from the biological process
underlying large, opportunistically collected biodiversity data, our
approach yields retrospective predictions on the occupancy of species
through time and space at high resolution and provides inference on
overall distributional trends, range dynamics, habitat preferences,
phenological patterns, detection patterns and observer heterogeneity.
Modelling the detection process accurately allows the model to weigh
visits based on their information content: outcomes of visits performed
by observers unlikely to detect or report the focal species, or visits
performed outside of the phenologically relevant period will not
strongly contribute to the occupancy estimation of the visited site.

By splitting up the input data into two data streams based on observer
expertise (or a proxy thereof) and treating them accordingly, both the
risk of false positive detection and information loss is minimised, making optimal use of the efforts of all observers. We believe this
approach is a practical alternative to using an explicit false positive
modelling approach, which we currently deem unfeasible for large
opportunistically sampled data due to computational limitations.

By stacking species-specific inferences, the present approach also
offers the opportunity to yield community-level insights, such as
estimates of species richness through space and time. Alternatively, a
flexible joint species distribution model accounting for imperfect
detection, as recently introduced by Fajgenblat \emph{et al.} (2025) may be
considered. Such multi-species extensions enable pooling of information
across species and yield valuable output for community ecology, such as
interspecific associations. We believe, however, that the proposed
single-species model will have an important place in modern biodiversity
research as large-scale, high resolution multi-species models tuned to
opportunistically sampled data can be prohibitively expensive to fit
without access to supercomputing facilities. Although fitting a single
multi-species model may require fewer computational resources than
fitting multiple single-species models for the same number of species,
the latter can be more convenient as it constitutes an embarrassingly
parallel task.

While our approach is tuned to the Observations.be portal, it can easily
be transferred to other data portals (e.g. iNaturalist, eBird, GBIF ...) and taxa using the provided script
with a generic pipeline. Notably, our model can readily be applied to a
wide variety of semi- and loosely structured datasets, as long as site
visits can explicitly or implicitly be derived from the data, and the
detection probability of the focal species can be assumed to be constant
among visits, conditionally on the included covariates. For instance,
nocturnal visits (e.g. for moth trapping, anuran and bat surveys \ldots)
should not be included in the analysis of diurnally active species, such
as butterflies, and vice-versa. The proposed model and the provided Stan
code can easily be extended to the specificities of specific projects or
study systems. For instance, the occupancy-related covariates can be
modelled through smoothing splines as an alternative to assuming linear
effects, or spatially-varying regression coefficients can be used for
larger study areas if deemed ecologically relevant (Finley 2010).
Recently, user-friendly R packages such as spOccupancy and flocker
became available, enabling practitioners to fit flexible
site-occupancies without bespoke probabilistic programming (Doser
\emph{et al.} 2022; Socolar \& Mills 2023). With some simplifications,
the proposed model can be fitted using these packages.

Despite the flexibility of the presented model, several challenges
intrinsic to biodiversity sampling in general or opportunistically
sampled data in particular still persist. First, the closure assumption,
a key assumption of site-occupancy models implying that the biological
process is constant throughout the primary study season (i.e. a full
year in our study), is likely to be violated when applying the model as
individuals can temporarily immigrate or emigrate from the focal site
within a single year, or as polyvoltine species' distribution varies
across generations within a single year. Such violations will confer
reduced detection probabilities (as the species is not always available
for detection), but they can be alleviated by terming the inferred
occupancy probabilities as ``probabilities of a site being used'' rather
than true occupancy probabilities. Second, spatiotemporal variation in
abundance might bias occupancy estimates as detection probabilities tend
to be positively related with abundance. For this reason, we chose not
to model temporal, spatial or habitat-dependent influences on detection
probabilities, as they might absorb abundance-induced effects on
detectability. Instead, we rather prefer the occupancy to be negatively
biased when abundance is low, and to be positively biased when abundance
is high. Third, preferential sampling, the phenomenon where sampling
effort is stochastically correlated with the biological process under
study, tends to be a dominant characteristic of opportunistic sampling
schemes, possibly biasing biological inferences. Preferential sampling
often leads to spatiotemporal sampling imbalance. Although this issue is
largely alleviated as our model intrinsically accounts for differences
in the number of visits across space and time, future research should be
devoted to more optimally correcting for preferential sampling within
our modelling framework, especially in a missing not at random (MNAR)
missingness scenario. Fourth, list length, the proxy for search effort
used in our approach, might be a weak predictor of true search effort,
as its upper bound varies seasonally, interannually and spatially.

Increasingly, biodiversity data portals facilitate and stimulate the
registration of formal metadata on visits by observers, for instance by
enabling the recording of their track alongside with the observations
they submit, yielding semi-structured data. From these registered
tracks, precise data on search effort can be derived and incorporated
into models. Since more efforts are required from the observer's side,
loosely structured data likely will still prevail in the future of
citizen science initiatives. Future developments will fill the gap between
loosely and semi-structured datasets, for instance by incorporating a
Bayesian imputation approach to address missing metadata and to
parsimoniously fuse both types of data in a single model.

In conclusion, our approach harnesses the strengths of large, loosely
structured opportunistically collected data, yielding valuable
information for ecologists, conservationists and policy-makers when
rigorously collected data is absent or too costly to collect. Furthermore, augmenting the value
of opportunistically collected data through a comprehensive statistical
analysis contributes to the acknowledgement and valorisation of the
efforts of many thousands of volunteer naturalists contributing data to
online biodiversity portals.

\newpage

\section*{References}
Belmont, J., Martino, S., Illian, J. \& Rue, H. (2024). spatiotemporal
occupancy models with INLA. \emph{Methods Ecol Evol}, 15, 2087--2100.

Betancourt, M. (2017). A Conceptual Introduction to Hamiltonian Monte
Carlo. \emph{ArXiv}.

Burns, F., Eaton, M.A., Hayhow, D.B., Outhwaite, C.L., Al Fulaij, N.,
August, T.A., \emph{et al.} (2018). An assessment of the state of nature
in the United Kingdom: A review of findings, methods and impact.
\emph{Ecol Indic}, 94, 226--236.

Carpenter, B., Gelman, A., Hoffman, M., Lee, D., Goodrich, B.,
Betancourt, M., \emph{et al.} (2017). Stan: A Probabilistic Programming
Language. \emph{Journal of Statistical Software, Articles}, 76, 1--32.

Doser, J.W., Finley, A.O., Kéry, M. \& Zipkin, E.F. (2022). spOccupancy:
An R package for single-species, multi-species, and integrated spatial
occupancy models. \emph{Methods Ecol Evol}, 13, 1670--1678.

European Environment Agency. (2022). \emph{CLCplus Backbone 2018 (raster
10 m), Europe, 3-yearly}.

Fajgenblat, M. \& Neyens, T. (2025). Temporal disaggregation through
interval-integrated B-splines for the integrated analysis of trapping
counts in ecology. bioRxiv, 2025.08. 07.669113.

Fajgenblat, M., Wijns, R., De Knijf, G., Stoks, R., Lemmens, P.,
Herremans, M., \emph{et al.} (2025). Leveraging Massive
Opportunistically Collected Datasets to Study Species Communities in
Space and Time. \emph{Ecol Lett}, 28.

Ferguson, P.F.B., Conroy, M.J. \& Hepinstall-Cymerman, J. (2015).
Occupancy models for data with false positive and false negative errors
and heterogeneity across sites and surveys. \emph{Methods Ecol Evol}, 6,
1395--1406.

Fitzpatrick, M.C., Preisser, E.L., Ellison, A.M. \& Elkinton, J.S.
(2009). Observer bias and the detection of low-density populations.
\emph{Ecological Applications}, 19, 1673--1679.

Hogeweg, L., Schermer, M., Pieterse, S., Roeke, T. \& Gerritsen, W.
(2019). Machine Learning Model for Identifying Dutch/Belgian
Biodiversity. \emph{Biodiversity Information Science and Standards}, 3.

Isaac, N.J.B., van Strien, A.J., August, T.A., de Zeeuw, M.P. \& Roy,
D.B. (2014). Statistics for citizen science: Extracting signals of
change from noisy ecological data. \emph{Methods Ecol Evol}, 5,
1052--1060.

Johnston, A., Fink, D., Hochachka, W.M. \& Kelling, S. (2018). Estimates
of observer expertise improve species distributions from citizen science
data. \emph{Methods Ecol Evol}, 9, 88--97.

Johnston, A., Moran, N., Musgrove, A., Fink, D. \& Baillie, S.R. (2020).
Estimating species distributions from spatially biased citizen science
data. \emph{Ecol Modell}, 422, 108927.

Joly, A., Bonnet, P., Goëau, H., Barbe, J., Selmi, S., Champ, J.,
\emph{et al.} (2016). A look inside the Pl@ntNet experience: The good,
the bias and the hope. \emph{Multimed Syst}, 22, 751--766.

Kellner, K.F. \& Swihart, R.K. (2014). Accounting for imperfect
detection in ecology: A quantitative review. \emph{PLoS One}, 9.

Kéry, M., Royle, J.A., Schmid, H., Schaub, M., Volet, B., Häfliger, G.,
\emph{et al.} (2010). Site-occupancy distribution modeling to correct
population-trend estimates derived from opportunistic observations.
\emph{Conservation Biology}, 24, 1388--1397.

Kéry, M. \& Schmidt, B. (2008). Imperfect detection and its consequences
for monitoring for conservation. \emph{Community Ecology}, 9, 207--216.

Knorr-Held, L. (2000). Bayesian modelling of inseparable space-time
variation in disease risk. \emph{Stat Med}, 19, 2555--2567.

MacKenzie, D.I., Nichols, J.D., Lachman, G.B., Droege, S., Royle, A.A.
\& Langtimm, C.A. (2002). Estimating site occupancy rates when detection
probabilities are less than one. \emph{Ecol}, 83, 2248--2255.

Maes, D. \& Van Dyck, H. (2001). Butterfly diversity loss in Flanders
(north Belgium): Europe's worst case scenario? \emph{Biol Conserv}, 99,
263--276.

Maes, D., Isaac, N.J.B., Harrower, C.A., Collen, B., van Strien, A.J. \&
Roy, D.B. (2015). The use of opportunistic data for IUCN Red List
assessments. \emph{Biological Journal of the Linnean Society}, 115,
690--706.

Maes, D., Vanreusel, W., Herremans, M., Vantieghem, P., Brosens, D.,
Gielen, K., \emph{et al.} (2016). A database on the distribution of
butterflies (Lepidoptera) in northern Belgium (Flanders and the Brussels
Capital Region). \emph{Zookeys}, 585, 143--156.

McClintock, B.T., Bailey, L.L., Pollock, K.H. \& Simons, T.R. (2010).
Unmodeled observation error induces bias when inferring patterns and
dynamics of species occurrence via aural detections. \emph{Ecology}, 91,
2446--2454.

Miller, D.A., Nichols, J.D., Mcclintock, B.T., Campbell Grant, E.H.,
Bailey, L.L. \& Weir, L.A. (2011). \emph{Improving occupancy estimation
when two types of observational error occur: non-detection and species
misidentification}. \emph{Ecology}.

Monnahan, C.C., Thorson, J.T. \& Branch, T.A. (2017). Faster estimation
of Bayesian models in ecology using Hamiltonian Monte Carlo.
\emph{Methods Ecol Evol}, 8, 339--348.

Monod, M., Blenkinsop, A., Brizzi, A., Chen, Y., Cardoso Correia
Perello, C., Jogarah, V., \emph{et al.} (2022). Regularised B-splines
Projected Gaussian Process Priors to Estimate Time-trends in
Age-specific COVID-19 Deaths. \emph{Bayesian Anal}, 1--31.

Neyens, T., Diggle, P.J., Faes, C., Beenaerts, N., Artois, T. \& Giorgi,
E. (2019). Mapping species richness using opportunistic samples: a case
study on ground-floor bryophyte species richness in the Belgian province
of Limburg. \emph{Sci Rep}, 9, 1--11.

NGI. (2022). \emph{Digital terrain model of Belgium (20 m resolution)}.

Outhwaite, C.L., Chandler, R.E., Powney, G.D., Collen, B., Gregory, R.D.
\& Isaac, N.J.B. (2018). Prior specification in Bayesian occupancy
modelling improves analysis of species occurence data. \emph{Ecol
Indic}, 93, 333--343.

Outhwaite, C.L., Gregory, R.D., Chandler, R.E., Collen, B. \& Isaac,
N.J.B. (2020). Complex long-term biodiversity change among
invertebrates, bryophytes and lichens. \emph{Nat Ecol Evol}, 4,
384--392.

Pocock, M.J.O., Logie, M.W., Isaac, N.J.B., Outhwaite, C.L. \& August,
T. (2019). Rapid assessment of the suitability of multi-species citizen
science datasets for occupancy trend analysis. \emph{bioRxiv}, 1--36.

R Core Team. (2020). R: A language and environment for statistical
computing. R Foundation for Statistical Computing, Vienna, Austria.

Rasmussen, C.Edward. \& Williams, C.K.I. (2006). \emph{Gaussian
processes for machine learning}. MIT Press.

Robinson, O.J., Ruiz-Gutierrez, V., Fink, D., Meese, R.J., Holyoak, M.
\& Cooch, E.G. (2018). Using citizen science data in integrated
population models to inform conservation decision-making.
\emph{bioRxiv}.

Royle, J.A. \& Link, W.A. (2006). Generalized site occupancy models
allowing for false positive and false negative errors. \emph{Ecology},
87, 835--841.

Ruiz-Gutierrez, V., Hooten, M.B. \& Campbell Grant, E.H. (2016).
Uncertainty in biological monitoring: a framework for data collection
and analysis to account for multiple sources of sampling bias.
\emph{Methods Ecol Evol}, 7, 900--909.

Shirey, V., Belitz, M.W., Barve, V. \& Guralnick, R. (2020). Closing
gaps but increasing bias in North American butterfly inventory
completeness.

Socolar, J.B. \& Mills, S.C. (2023). Introducing flocker: an R package
for flexible occupancy modeling via brms and Stan.

Soroye, P., Ahmed, N. \& Kerr, J.T. (2018). Opportunistic citizen
science data transform understanding of species distributions,
phenology, and diversity gradients for global change research.
\emph{Glob Chang Biol}, 24, 5281--5291.

Van Strien, A.J., Gmelig, A.W., Herder, J.E., Hollander, H., Kalkman,
V.J., Poot, M.J.M., \emph{et al.} (2016). Modest recovery of
biodiversity in a western European country: The Living Planet Index for
the Netherlands. \emph{Biol Conserv}, 200, 44--50.

Van Strien, A.J., van Swaay, C.A.M. \& Kéry, M. (2011). Metapopulation
dynamics in the butterfly Hipparchia semele changed decades before
occupancy declined in The Netherlands. \emph{Ecol Appl}, 21, 2510--2520.

Van Strien, A.J., Van Swaay, C.A.M., van Strien-van Liempt, W.T.F.H.,
Poot, M.J.M. \& WallisDeVries, M.F. (2019). Over a century of data
reveal more than 80\% decline in butterflies in the Netherlands.
\emph{Biol Conserv}, 234, 116--122.

Van Strien, A.J., Van Swaay, C.A.M. \& Termaat, T. (2013). Opportunistic
citizen science data of animal species produce reliable estimates of
distribution trends if analysed with occupancy models. \emph{Journal of
Applied Ecology}, 50, 1450--1458.

Van Strien, A.J., Van Swaay, C.A.M. \& Termaat, T. (2013). Opportunistic
citizen science data of animal species produce reliable estimates of
distribution trends if analysed with occupancy models. \emph{J Appl
Ecol}, 50, 1450--1458.

Termaat, T., Van Grunsven, R.H.A., Plate, C.L. \& Van Strien, A.J.
(2015). Strong recovery of dragonflies in recent decades in The
Netherlands. \emph{Freshwater Science}, 34, 1094--1104.

Termaat, T., van Strien, A.J., van Grunsven, R.H.A., De Knijf, G.,
Bjelke, U., Burbach, K., \emph{et al.} (2019). Distribution trends of
European dragonflies under climate change. \emph{Divers Distrib}, 25,
936--950.

Van Eupen, C., Maes, D., Herremans, M., Swinnen, K.R.R., Somers, B. \& Luca, S. (2021). The impact of data quality filtering of opportunistic citizen science data on species distribution model performance. Ecological Modelling 444: 109453.

Vanreusel, W., Herremans, M., Vantieghem, P., Gielen, K., Desmet, P. \&
Swinnen, K. (2019). \emph{Waarnemingen.be - Butterfly occurrences in
Flanders and the Brussels Capital Region, Belgium. Version 1.9.
Natuurpunt. Occurrence dataset}. \emph{GBIF.org}.

Vantieghem, P., Maes, D., Kaiser, A. \& Merckx, T. (2017). Quality of
citizen science data and its consequences for the conservation of
skipper butterflies (Hesperiidae) in Flanders (northern Belgium).
\emph{J Insect Conserv}, 21, 451--463.

Vehtari, A., Gelman, A., Simpson, D., Carpenter, B. \& Bürkner, P.C.
(2019). Rank-normalization, folding, and localization: An improved R for
assessing convergence of MCMC. \emph{ArXiv}, 1--27.

Warren, M.S., Maes, D., van Swaay, C.A.M., Goffart, P., Van Dyck, H.,
Bourn, N.A.D., Wynhoff, I., Hoare, D.J. \& Ellis, S. \emph{et al.} (2021). The decline of butterflies in
Europe: problems, significance, and possible solutions.
\emph{Proceedings of the National Academy of Science of the United
States of America}, 118(2): e2002551117.

Yackulic, C.B., Dodrill, M., Dzul, M., Sanderlin, J.S. \& Reid, J.A.
(2020). A need for speed in Bayesian population models: a practical
guide to marginalizing and recovering discrete latent states.
\emph{Ecological Applications}, 30, 1--19.

\end{document}